\documentclass[8pt]{revtex4}
\usepackage{amssymb}
\usepackage{latexsym}
\usepackage{epsfig}
\begin{document}

\title{Tachyonic  $\delta$-Tsallis entropy of  a thermal tachyonic BIon}

\author{ Aroonkumar Beesham$^{1}$\footnote{beeshama@unizulu.ac.za}, Alireza Sepehri$^{1,2}$\footnote{alireza.sepehri3@gmail.com}}
\affiliation{$^{1}$Department of Mathematical Sciences, University of Zululand, Private Bag X1001, Kwa-Dlangezwa 3886, South Africa\\$^{2}$Research Institute for Astronomy and Astrophysics of
Maragha (RIAAM), P.O. Box 55134-441, Maragha, Iran }

\begin{abstract}
When a brane and an anti-brane come close to each other, the tachyonic potential between them increases and a tachyon wormhole is formed. This configuration, which consists of two branes and a tachyonic wormhole, is called a thermal tachyonic BIon. By considering the thermodynamic behaviour of this system, one  finds that its entropy has the same form as that of the Tsallis one. By decreasing the separation between the branes,  the tachyonic potential increases, and the entropy grows.
\vspace{5mm}\noindent\vspace{0.8mm}\newline Keywords: Tachyon, Tsallis entropy, BIon, Area
\end{abstract}

\maketitle
\section{Introduction}

Recently,  Tsallis  and  Cirto  have  proposed  that  the  entropy  of  a  gravitational  system  such  as  a  black  hole  could  be  generalized
to  the  non-additive  entropy,  which  is  given  by  $S = \gamma A^{\beta}$  ,
where  A  is  the  horizon  area \cite{ss1}. There  has  been  lots  of  discussion  on  this  topic  so  far.  For  example,  some  authors  have  investigated  the  limited  behaviour  of  the  evolution  of  the Tsallis  entropy  in  self-gravitating  systems.  They  have  argued  that
the  Tsallis  entropy  generally  exhibits  a  bounded  property  in  a self-gravitating  system.  This  shows  the  existence  of a global  maximum  of  the Tsallis  entropy \cite{ss2}. Some  other  authors  have
proposed  a  coherence  quantiﬁer  in  terms  of  the  Tsallis  relative   entropy,  which  lays  the  foundation  for  non-extensive  thermo-statistics  and  plays  the  same  role  as  the  standard  logarithmic  entropy  does  in    information  theory \cite{ss3,ss4,ss5}. In
  another  consideration,  some authors  have  derived  the entropic-force
  terms  from  a  generalized  black-hole  entropy  proposed  by
  Tsallis  and   Cirto  in  order  to  examine   entropic
cosmology.  Unlike  the  Bekenstein  entropy,  which  is  proportional  to  area, generalized  entropy  is  proportional  to  volume  because  of  appropriate  nonadditive  generalizations \cite{ss6}. 

In  another  work,  the  relation  between  the Tsallis  entropy  and  the  exchange  of  energy  between  the  bulk  (the  universe)  and  the boundary (the horizon of the universe) has been considered \cite{ss7}.  In another investigation, Using the Tsallis entropy, the evolution  of the universe in entropic cosmologies has been studied. In this model, the authors have considered an extended entropic-
force model that includes a Hubble parameter (H) term and a
constant term in entropic-force terms. The H term is derived
from a volume entropy, whereas the constant term is derived
from an entropy proportional to the square of an area \cite{ss8}. In another research, the evolution of the Tsallis entropy during
non-adiabatic processes like the accelerated expansion of the late uni-
verse has been considered \cite{ss9}. In addition, the application
of this entropy in other aspects of cosmology and physics
has been investigated \cite{ss10,ss11}. And ﬁnally,
employing the modiﬁed entropy-area relation suggested by
Tsallis and Cirto and the holographic hypothesis, a new holo-
graphic dark energy (HDE) model was proposed \cite{ss12}.

In this paper, we will show that the Tsallis entropy could be produced by a tachyonic potential in a brane-anti-brane system. This potential produces a tachyonic wormhole between branes and leads to the formation of a BIon.  A BIon is a configuration which is formed from a brane, an anti-brane and a wormhole which connects them \cite{ss13,ss14,ss15}. By increasing the tachyonic potential, thermodynamically, the behaviour of this BIon changes and its Tsallis entropy grows.

 The outline of the
paper is as  follows.  In section II, we will  consider the dependency of the area of the  BIon on tachyon fields. In section
III, we will consider dependency on tachyon fields.

\section{ Dependency of area of a BIon on tachyons }\label{o1}

In this section, we will show that tachyonic potential produces a wormhole between a brane and an anti-brane. This system which is formed by joining a brane, an anti-brane and a wormhole called a BIon. We obtain the area of this tachyonic BIon.

\begin{figure*}[thbp]
	\begin{center}
		\begin{tabular}{rl}
			\includegraphics[width=8cm]{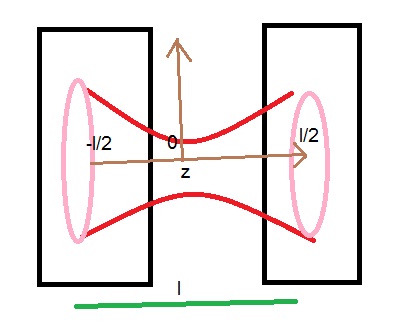}
		\end{tabular}
	\end{center}
	\caption{A set of
$D3$-$\overline{D3}$-brane pairs  which
are placed at points $z_{1} = l/2$ and $z_{2} = -l/2$ respectively
so that the separation between the brane and antibrane is $l$.  }
\end{figure*}

\begin{figure*}[thbp]
	\begin{center}
		\begin{tabular}{rl}
			\includegraphics[width=14cm]{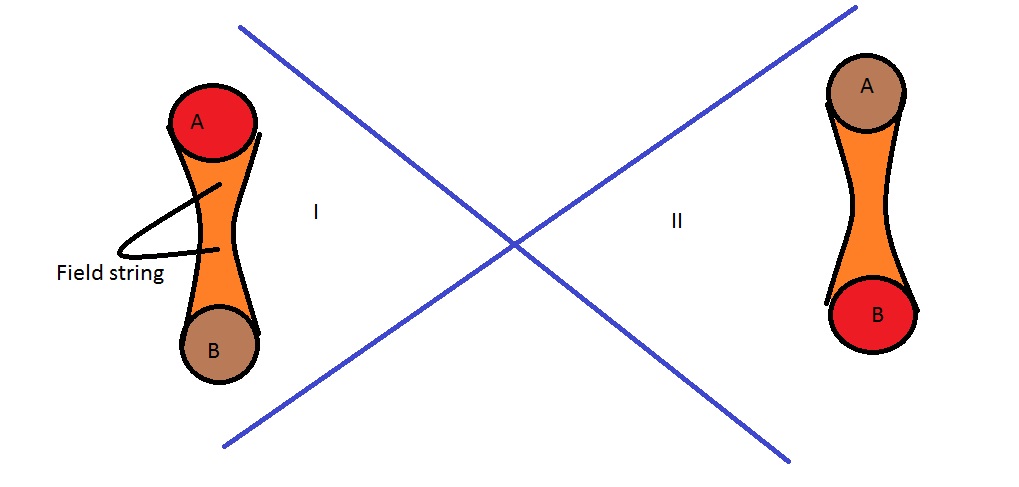}
		\end{tabular}
	\end{center}
	\caption{ BIon in a rindler space-time}
\end{figure*}

To obtain the tachyonic potential and construct  a tachyonic black hole in this theory, we consider a set of
$D3$-$\overline{D3}$-brane pairs  which
are placed at points $z_{1} = l/2$ and $z_{2} = -l/2$, respectively, 
so that the separation between the brane and antibrane is $l$.  Let $z$ be a transverse coordinate to the branes and $\sigma$
be the radius on the world-volume (See figure 1).  The induced metric on the
brane is:
\begin{eqnarray}
\gamma_{ab}d\sigma^{a}d\sigma^{b} = -d\tau^{2} + (1 +
z'(\sigma)^{2})d\sigma^{2} + \sigma^{2}(d\theta^{2} +
\sin^{2}\theta d\phi^{2}) \label{Q5}
\end{eqnarray}

 For
the simple case of a single $D3$-$\overline{D3}$-brane pair with
open string tachyon, the action is \cite{ss16,ss17,ss18,ss19}:
 \begin{eqnarray}
&& S_{tot-extra}=-\tau_{3}\int d^{9}\sigma \sum_{i=1}^{2}
V(TA,l)e^{-\phi}(\sqrt{-det A_{i}})\nonumber \\&&
(A_{i})_{ab}=\left(g_{MN}-\frac{TA^{2}l^{2}}{Q}g_{Mz}g_{zN}\right)\partial_{a}x^{M}_{i}\partial_{b}x^{M}_{i}
+F^{i}_{ab}+\frac{1}{2Q}((D_{a}TA)(D_{b}TA)^{\ast}+(D_{a}TA)^{\ast}(D_{b}TA))\nonumber
\\&&
+il(g_{az}+\partial_{a}z_{i}g_{zz})(TA(D_{b}TA)^{\ast}-TA^{\ast}(D_{b}TA))+
il(TA(D_{a}TA)^{\ast}-TA^{\ast}(D_{a}TA))(g_{bz}+\partial_{b}z_{i}g_{zz}),
\label{Q27}
\end{eqnarray}
where
  \begin{eqnarray}
&& Q=1+TA^{2}l^{2}g_{zz}, \nonumber \\&&
D_{a}TA=\partial_{a}TA-i(A_{2,a}-A_{1,a})TA,
V(TA,l)=g_{s}V(TA)\sqrt{Q}, \nonumber \\&& e^{\phi}=g_{s}( 1 +
\frac{R^{4}}{z^{4}} )^{-\frac{1}{2}}, \label{Q28}
\end{eqnarray}
The quantities $\phi$, $A_{2,a}$ and $F^{i}_{ab}$ are the dilaton field, gauge
fields and field strengths on the world-volume of the non-BPS
brane, respectively;  $TA$ is the tachyon field, $\tau_{3}$ is the brane tension
and $V(TA)$ is the tachyon potential. The indices $a,b$ denote the
tangent directions of the $D$-branes, while the indices $M,N$ run over the
background ten-dimensional space-time directions. The $Dp$-brane and
the anti-$Dp$-brane are labeled by $i$ = 1 and 2, respectively. Then
the separation between these $D$-branes is defined by $z_{2} - z_{1}
= l$. Also, in writing the above action, we are using the convention
$2\pi\acute{\alpha}=1$.

Let us consider the action of a D3-brane and for simplicity, we consider only a $\sigma$
dependence of the tachyon field $TA$,  and set the
gauge fields to zero. In this case, the action (\ref{Q27}) in the
region that  $r> R$ and $TA'\sim constant$ simplifies to
  \begin{eqnarray}
S_{D3} \simeq-\frac{\tau_{3}}{g_{s}}\int dt \int d\sigma \sigma^{2}
V(TA)(\sqrt{D_{1,TA}}+\sqrt{D_{2,TA}}), \label{Q29}
\end{eqnarray}
where $D_{1,TA} = D_{2,TA}\equiv D_{TA}$,
${\displaystyle V_{3}=\frac{4\pi^{2}}{3}}$ is the volume of a unit sphere $S^{3}$ and
 \begin{eqnarray}
D_{TA} = 1 + \frac{l'(\sigma)^{2}}{4}+ TA^{2}l^{2}, \label{Q30}
\end{eqnarray}
where the prime  denotes a derivative with respect to
$\sigma$. A useful potential  that can be used is
\cite{ss16}:
 \begin{eqnarray}
V(TA)=\frac{\tau_{3}}{\cosh\sqrt{\pi}TA}. \label{Q31}
\end{eqnarray}
The energy momentum tensor is obtained from the action by calculating
its functional derivative with respect to the  ten-dimensional background
metric $g_{MN}$. The  variation  is ${\displaystyle T^{MN} =
\frac{2}{\sqrt{-det g}}\frac{\delta S}{\delta g_{MN}}}$. We get
\cite{ss15},
 \begin{eqnarray}
&& T^{00}_{i,brane}=V(TA)\sqrt{D_{TA}},  \label{Q32}
\end{eqnarray}

After doing some calculations and using  some approximations, we obtain:

\begin{eqnarray}
&& T^{00}_{i,brane}= \tau_{3} + V_{brane},  \label{moment1}
\end{eqnarray}
where

\begin{eqnarray}
&&   V_{brane}= \tau_{3}[\frac{\sqrt{\pi}TA}{2}][1+ e^{-2\sqrt{\pi}TA}]^{-1}\times \nonumber
\\&& [\frac{l'(\sigma)^{2}}{4}+ TA^{2}l^{2}] \label{pot2}
\end{eqnarray}
This potential depends on the tachyon and the separation distance between the two branes. To obtain the dependence of these parameters in terms of time, we should regard the effects of other branes. We will show that when branes come close to each other, the tachons produce a wormhole which connects the branes and transmits energy from the extra dimensions into our black hole. 

Until now, we have considered  that the tachyon field
grows slowly ($TA \sim t^{4}/t^{3}= t$), and we ignored
${\displaystyle TA'=\frac{\partial TA}{\partial \sigma}}$ and
${\displaystyle \dot{TA}=\frac{\partial TA}{\partial t}}$ in our calculations. In
this section, we show that, with the decrease of the distance 
 between the brane and antibrane black holes, the tachyon field grows very fast and
$TA'$ and $\dot{TA}$ cannot be discarded. These dynamics lead to   the formation of a  new wormhole.
 In this stage, the black hole evolves from non-phantom phase to a new
phantom phase and consequently, the phantom-dominated era of the
black hole accelerates and ends up in the  big-rip singularity. In this
case, the action (\ref{Q27}) is given by the following Lagrangian $L$:
  \begin{eqnarray}
L \simeq-\frac{\tau_{3}}{g_{s}} \int d\sigma \sigma^{2}
V(TA)(\sqrt{D_{1,TA}}+\sqrt{D_{2,TA}}), \label{Q35}
\end{eqnarray}
where
 \begin{eqnarray}
D_{1,TA} = D_{2,TA}\equiv D_{TA} = 1 + \frac{l'(\sigma)^{2}}{4}+ 
\dot{TA}^{2} -  TA'^{2}+ TA^{2}l^{2}, \label{Q36}
\end{eqnarray}
and we assume that $TA l\ll TA'$. Now, we  study the Hamiltonian
corresponding to the above Lagrangian.
 In order to derive such a Hamiltonian,  we need the canonical momentum density ${\displaystyle \Pi =
\frac{\partial L}{\partial \dot{TA}}}$ associated with the tachyon, i.e., 
 \begin{eqnarray}
\Pi = \frac{V(TA)\dot{TA}}{ \sqrt{1 + \frac{l'(\sigma)^{2}}{4}+
\dot{TA}^{2} -  TA'^{2}}}, \label{Q37}
\end{eqnarray}
so that the Hamiltonian can be obtained as:
\begin{eqnarray}
H_{DBI} = 4\pi\int d\sigma  \sigma^{2} \Pi \dot{TA} - L.
 \label{Q38}
\end{eqnarray}
By choosing $\dot{TA} = 2 TA'$, this gives:
\begin{eqnarray}
H_{DBI} = 4\pi\int d\sigma \sigma^{2} \left[\Pi
(\dot{TA}-\frac{1}{2}TA')\right] + \frac{1}{2}TA\partial_{\sigma}(\Pi
\sigma^{2}) - L
 \label{Q39}
\end{eqnarray}
In this equation, we have, in the second step, integrated by parts
the term proportional to $\dot{TA}$, indicating that the tachyon can
be studied as a Lagrange multiplier imposing the constraint
$\partial_{\sigma}(\Pi \sigma^{2}V(TA))=0$ on the canonical
momentum. Solving this equation yields:
\begin{eqnarray}
\Pi =\frac{\beta}{4\pi \sigma^{2}},
 \label{Q40}
\end{eqnarray}
where $\beta$ is a constant.

Using equations (\ref{Q37} and \ref{Q40}),  and assuming ($l' << TA'$), we can obtain $\sigma$ in terms of tachyons:

\begin{eqnarray}
\sigma =[\frac{ [4\pi \sqrt{1 +
\dot{TA}^{2} -  TA'^{2}}]}{\beta V(TA)\dot{TA}}]^{\frac{1}{2}}
 \label{Qq40}
\end{eqnarray}

Taking the derivative of the above equation with respect to time, we  obtain the acceleration of system:

\begin{eqnarray}
a =\frac{d^{2}}{d t^{2}}\sigma =\frac{d^{2}}{d t^{2}}[\frac{ [4\pi \sqrt{1 +
\dot{TA}^{2} -  TA'^{2}}]}{\beta V(TA)\dot{TA}}]^{\frac{1}{2}}
 \label{Qqc40}
\end{eqnarray}

The above equation shows that acceleration of the BIon has a direct relation with tachyonic fields which live on it. This acceleration leads to the emergence of a Rindler space-time (See figure 2). In these conditions,  the relation between  the world volume coordinates of the BIon ($\tau, \sigma $) and the coordinates of
Minkowski space-time ($t, r$) are \cite{ss20};

\begin{eqnarray}
&& at= e^{a\sigma} \sinh(a\tau) \quad ar=e^{a\sigma} \cosh(a\tau) \quad \text{In Region I} \nonumber\\&& at= - e^{-a\sigma} \sinh(a\tau) \quad ar =  e^{-a\sigma} \cosh(a\tau) \quad \text{In Region II}
\label{a9}
\end{eqnarray}

Now, we can obtain the metric of a thermal BIon in non-flat space-time. Replacing the  acceleration by tachyonic fields in equation (\ref{Qqc40}), we can rewrite equation (\ref{a9}) as:

\begin{eqnarray}
&&  [\frac{d^{2}}{d t^{2}}[\frac{ [4\pi \sqrt{1 +
\dot{TA}^{2} -  TA'^{2}}]}{\beta V(TA)\dot{TA}}]^{\frac{1}{2}}]t= e^{ [\frac{d^{2}}{d t^{2}}[\frac{ [4\pi \sqrt{1 +
\dot{TA}^{2} -  TA'^{2}}]}{\beta V(TA)\dot{TA}}]^{\frac{1}{2}}]\sigma} \sinh(a\tau) \nonumber\\&&  [\frac{d^{2}}{d t^{2}} [\frac{d^{2}}{d t^{2}}[\frac{ [4\pi \sqrt{1 +
\dot{TA}^{2} -  TA'^{2}}]}{\beta V(TA)\dot{TA}}]^{\frac{1}{2}}]r=e^{ [\frac{d^{2}}{d t^{2}}[\frac{ [4\pi \sqrt{1 +
\dot{TA}^{2} -  TA'^{2}}]}{\beta V(TA)\dot{TA}}]^{\frac{1}{2}}]\sigma} \cosh(a\tau) \quad \text{In Region I} \nonumber\\&&  [\frac{d^{2}}{d t^{2}}[\frac{ [4\pi \sqrt{1 +
\dot{TA}^{2} -  TA'^{2}}]}{\beta V(TA)\dot{TA}}]^{\frac{1}{2}}]t= - e^{- [\frac{d^{2}}{d t^{2}}[\frac{ [4\pi \sqrt{1 +
\dot{TA}^{2} -  TA'^{2}}]}{\beta V(TA)\dot{TA}}]^{\frac{1}{2}}]\sigma} \sinh(a\tau) \nonumber\\&&  [\frac{d^{2}}{d t^{2}}[\frac{ [4\pi \sqrt{1 +
\dot{TA}^{2} -  TA'^{2}}]}{\beta V(TA)\dot{TA}}]^{\frac{1}{2}}]r =  e^{-[\frac{d^{2}}{d t^{2}}[\frac{ [4\pi \sqrt{1 +
\dot{TA}^{2} -  TA'^{2}}]}{\beta V(TA)\dot{TA}}]^{\frac{1}{2}}]\sigma} \cosh(a\tau) \quad \text{In Region II}
\label{a10}
\end{eqnarray}

The above equation shows that tachyonic fields change the coordinates of space-time, lead to acceleration and produce two different regions in a new Rindler space-time. Thus, the metric changes, and a new metric in regions I and II emerge.

Substituting equation (\ref{a10}) in equation (\ref{a9}), we obtain:

\begin{eqnarray}
&& ds^{2}_{I,A,thermal}= D_{I-A}^{\frac{1}{2}} H_{I-A}^{-\frac{1}{2}}f_{I-A}\times  \nonumber\\&& \Big(e^{2[\frac{d^{2}}{d t^{2}}[\frac{ [4\pi \sqrt{1 +
\dot{TA}^{2} -  TA'^{2}}]}{\beta V(TA)\dot{TA}}]^{\frac{1}{2}}]\sigma} + \frac{1}{\sinh^{2}( [\frac{d^{2}}{d t^{2}}[\frac{ [4\pi \sqrt{1 +
\dot{TA}^{2} -  TA'^{2}}]}{\beta V(TA)\dot{TA}}]^{\frac{1}{2}}]\tau)}(\frac{dz}{d\tau})^{2} \Big)d\tau^{2} -  \nonumber\\&& D_{I-A}^{-\frac{1}{2}} H_{I-A}^{\frac{1}{2}}f_{I-A}^{-1}\times  \nonumber\\&&\Big(e^{2 [\frac{d^{2}}{d t^{2}}[\frac{ [4\pi \sqrt{1 +
\dot{TA}^{2} -  TA'^{2}}]}{\beta V(TA)\dot{TA}}]^{\frac{1}{2}}]\sigma}+ \frac{1}{\cosh^{2}( [\frac{d^{2}}{d t^{2}}[\frac{ [4\pi \sqrt{1 +
\dot{TA}^{2} -  TA'^{2}}]}{\beta V(TA)\dot{TA}}]^{\frac{1}{2}}]\tau)}(\frac{dz}{d\sigma})^{2} \Big) d\sigma^{2} +  \nonumber\\&& \frac{1}{\sinh([\frac{d^{2}}{d t^{2}}[\frac{ [4\pi \sqrt{1 +
\dot{TA}^{2} -  TA'^{2}}]}{\beta V(TA)\dot{TA}}]^{\frac{1}{2}}]\tau)\cosh( [\frac{d^{2}}{d t^{2}}[\frac{ [4\pi \sqrt{1 +
\dot{TA}^{2} -  TA'^{2}}]}{\beta V(TA)\dot{TA}}]^{\frac{1}{2}}]\tau)}(\frac{dz}{d\tau }\frac{dz}{d\sigma})d\tau d\sigma +  \nonumber\\&&  D_{I-A}^{-\frac{1}{2}} H_{I-A}^{\frac{1}{2}}\Big(\frac{1}{[\frac{d^{2}}{d t^{2}}[\frac{ [4\pi \sqrt{1 +
\dot{TA}^{2} -  TA'^{2}}]}{\beta V(TA)\dot{TA}}]^{\frac{1}{2}}]}e^{ [\frac{d^{2}}{d t^{2}}[\frac{ [4\pi \sqrt{1 +
\dot{TA}^{2} -  TA'^{2}}]}{\beta V(TA)\dot{TA}}]^{\frac{1}{2}}]\sigma} \cosh([\frac{d^{2}}{d t^{2}}[\frac{ [4\pi \sqrt{1 +
\dot{TA}^{2} -  TA'^{2}}]}{\beta V(TA)\dot{TA}}]^{\frac{1}{2}}]\tau)\Big)^{2}\times  \nonumber\\&&\Big(d\theta^{2} + sin^{2}\theta d\phi^{2}\Big)  + \nonumber\\&& D_{I-A}^{-\frac{1}{2}} H_{I-A}^{-\frac{1}{2}} \sum_{i=1}^{5}dx_{i}^{2}
\label{a17}
\end{eqnarray}

\begin{eqnarray}
&& ds^{2}_{II,A,thermal}= D_{II-A}^{\frac{1}{2}} H_{II-A}^{-\frac{1}{2}}f_{II-A}\times  \nonumber\\&&\Big(e^{-2 [\frac{d^{2}}{d t^{2}}[\frac{ [4\pi \sqrt{1 +
\dot{TA}^{2} -  TA'^{2}}]}{\beta V(TA)\dot{TA}}]^{\frac{1}{2}}]\sigma} + \frac{1}{\sinh^{2}([\frac{d^{2}}{d t^{2}}[\frac{ [4\pi \sqrt{1 +
\dot{TA}^{2} -  TA'^{2}}]}{\beta V(TA)\dot{TA}}]^{\frac{1}{2}}]\tau)}(\frac{dz}{d\tau})^{2} \Big)d\tau^{2} - \nonumber\\&& D_{II-A}^{-\frac{1}{2}} H_{II-A}^{\frac{1}{2}}f_{II-A}^{-1} \times  \nonumber\\&&\Big(e^{-2 [\frac{d^{2}}{d t^{2}}[\frac{ [4\pi \sqrt{1 +
\dot{TA}^{2} -  TA'^{2}}]}{\beta V(TA)\dot{TA}}]^{\frac{1}{2}}]\sigma}+ \frac{1}{\cosh^{2}( [\frac{d^{2}}{d t^{2}}[\frac{ [4\pi \sqrt{1 +
\dot{TA}^{2} -  TA'^{2}}]}{\beta V(TA)\dot{TA}}]^{\frac{1}{2}}]\tau)}(\frac{dz}{d\sigma})^{2} \Big) d\sigma^{2} - \nonumber\\&& \frac{1}{\sinh( [\frac{d^{2}}{d t^{2}}[\frac{ [4\pi \sqrt{1 +
\dot{TA}^{2} -  TA'^{2}}]}{\beta V(TA)\dot{TA}}]^{\frac{1}{2}}]\tau)\cosh(a\tau)}(\frac{dz}{d\tau }\frac{dz}{d\sigma})d\tau d\sigma + \nonumber\\&&  D_{II-A}^{-\frac{1}{2}} H_{II-A}^{\frac{1}{2}}\Big(\frac{1}{ [\frac{d^{2}}{d t^{2}}[\frac{ [4\pi \sqrt{1 +
\dot{TA}^{2} -  TA'^{2}}]}{\beta V(TA)\dot{TA}}]^{\frac{1}{2}}]\sigma} \cosh( [\frac{d^{2}}{d t^{2}}[\frac{ [4\pi \sqrt{1 +
\dot{TA}^{2} -  TA'^{2}}]}{\beta V(TA)\dot{TA}}]^{\frac{1}{2}}]\tau)\Big)^{2}\Big(d\theta^{2} + sin^{2}\theta d\phi^{2}\Big)  +  \nonumber\\&& D_{II-A}^{-\frac{1}{2}} H_{II-A}^{-\frac{1}{2}}\sum_{i=1}^{5}dx_{i}^{2}
\label{a18}
\end{eqnarray}

where

\begin{eqnarray}
&& f_{I-A} = 1-\frac{\Big(e^{[\frac{d^{2}}{d t^{2}}[\frac{ [4\pi \sqrt{1 +
\dot{TA}^{2} -  TA'^{2}}]}{\beta V(TA)\dot{TA}}]^{\frac{1}{2}}]\sigma_{0}} \cosh( [\frac{d^{2}}{d t^{2}}[\frac{ [4\pi \sqrt{1 +
\dot{TA}^{2} -  TA'^{2}}]}{\beta V(TA)\dot{TA}}]^{\frac{1}{2}}]\tau_{0})\Big)^{4}}{\Big(e^{ [\frac{d^{2}}{d t^{2}}[\frac{ [4\pi \sqrt{1 +
\dot{TA}^{2} -  TA'^{2}}]}{\beta V(TA)\dot{TA}}]^{\frac{1}{2}}]\sigma} \cosh([\frac{d^{2}}{d t^{2}}[\frac{ [4\pi \sqrt{1 +
\dot{TA}^{2} -  TA'^{2}}]}{\beta V(TA)\dot{TA}}]^{\frac{1}{2}}]\tau)\Big)^{4}}  \nonumber\\&& H_{I-A} = 1+\frac{\Big(e^{ [\frac{d^{2}}{d t^{2}}[\frac{ [4\pi \sqrt{1 +
\dot{TA}^{2} -  TA'^{2}}]}{\beta V(TA)\dot{TA}}]^{\frac{1}{2}}]\sigma_{0}} \cosh( [\frac{d^{2}}{d t^{2}}[\frac{ [4\pi \sqrt{1 +
\dot{TA}^{2} -  TA'^{2}}]}{\beta V(TA)\dot{TA}}]^{\frac{1}{2}}]\tau_{0})\Big)^{4}\sinh^{2}\alpha_{I-A}}{\Big(e^{ [\frac{d^{2}}{d t^{2}}[\frac{ [4\pi \sqrt{1 +
\dot{TA}^{2} -  TA'^{2}}]}{\beta V(TA)\dot{TA}}]^{\frac{1}{2}}]\sigma} \cosh( [\frac{d^{2}}{d t^{2}}[\frac{ [4\pi \sqrt{1 +
\dot{TA}^{2} -  TA'^{2}}]}{\beta V(TA)\dot{TA}}]^{\frac{1}{2}}]\tau)\Big)^{4}}\nonumber\\&& D_{I-A} = \cos^{2}\epsilon_{I-A} + \sin^{2}\epsilon_{I-A} H_{I-A}^{-1}
\label{a19}
\end{eqnarray}

\begin{eqnarray}
&& f_{II-A} = 1-\frac{\Big(e^{-[\frac{d^{2}}{d t^{2}}[\frac{ [4\pi \sqrt{1 +
\dot{TA}^{2} -  TA'^{2}}]}{\beta V(TA)\dot{TA}}]^{\frac{1}{2}}]\sigma_{0}} \cosh([\frac{d^{2}}{d t^{2}}[\frac{ [4\pi \sqrt{1 +
\dot{TA}^{2} -  TA'^{2}}]}{\beta V(TA)\dot{TA}}]^{\frac{1}{2}}]\tau_{0})\Big)^{4}}{\Big(e^{-[\frac{d^{2}}{d t^{2}}[\frac{ [4\pi \sqrt{1 +
\dot{TA}^{2} -  TA'^{2}}]}{\beta V(TA)\dot{TA}}]^{\frac{1}{2}}]\sigma} \cosh([\frac{d^{2}}{d t^{2}}[\frac{ [4\pi \sqrt{1 +
\dot{TA}^{2} -  TA'^{2}}]}{\beta V(TA)\dot{TA}}]^{\frac{1}{2}}]\tau)\Big)^{4}}  \nonumber\\&& H_{II-A} = 1+\frac{\Big(e^{[\frac{d^{2}}{d t^{2}}[\frac{ [4\pi \sqrt{1 +
\dot{TA}^{2} -  TA'^{2}}]}{\beta V(TA)\dot{TA}}]^{\frac{1}{2}}]\sigma_{0}} \cosh([\frac{d^{2}}{d t^{2}}[\frac{ [4\pi \sqrt{1 +
\dot{TA}^{2} -  TA'^{2}}]}{\beta V(TA)\dot{TA}}]^{\frac{1}{2}}]\tau_{0})\Big)^{4}\sinh^{2}\alpha_{II-A}}{\Big(e^{[\frac{d^{2}}{d t^{2}}[\frac{ [4\pi \sqrt{1 +
\dot{TA}^{2} -  TA'^{2}}]}{\beta V(TA)\dot{TA}}]^{\frac{1}{2}}]\sigma} \cosh([\frac{d^{2}}{d t^{2}}[\frac{ [4\pi \sqrt{1 +
\dot{TA}^{2} -  TA'^{2}}]}{\beta V(TA)\dot{TA}}]^{\frac{1}{2}}]\tau)\Big)^{4}}\nonumber\\&& D_{II-A} = \cos^{2}\epsilon_{II-A} + \sin^{2}\epsilon_{II-A} H_{II-A}^{-1}
\label{a20}
\end{eqnarray}

and

\begin{eqnarray}
&& \cosh^{2} \alpha_{I-A} = \frac{3}{2}\frac{\cos\frac{\delta_{I-A}}{3} + \sqrt{3}\cos\frac{\delta_{I-A}}{3}}{\cos\delta_{I-A}}\nonumber\\&& \cos\epsilon_{I-A} = \frac{1}{\sqrt{1 + \frac{K^{2}}{\Big( [\frac{d^{2}}{d t^{2}}[\frac{ [4\pi \sqrt{1 +
\dot{TA}^{2} -  TA'^{2}}]}{\beta V(TA)\dot{TA}}]^{\frac{1}{2}}]^{-1}e^{-[\frac{d^{2}}{d t^{2}}[\frac{ [4\pi \sqrt{1 +
\dot{TA}^{2} -  TA'^{2}}]}{\beta V(TA)\dot{TA}}]^{\frac{1}{2}}]\sigma} \cosh([\frac{d^{2}}{d t^{2}}[\frac{ [4\pi \sqrt{1 +
\dot{TA}^{2} -  TA'^{2}}]}{\beta V(TA)\dot{TA}}]^{\frac{1}{2}}]\tau)\Big)^{4}}}}
\label{a21}
\end{eqnarray}

\begin{eqnarray}
&& \cosh^{2} \alpha_{II-A} = \frac{3}{2}\frac{\cos\frac{\delta_{II-A}}{3} + \sqrt{3}\cos\frac{\delta_{II-A}}{3}}{\cos\delta_{II-A}}\nonumber\\&& \cos\epsilon_{II-A} = \frac{1}{\sqrt{1 + \frac{K^{2}}{\Big( [\frac{d^{2}}{d t^{2}}[\frac{ [4\pi \sqrt{1 +
\dot{TA}^{2} -  TA'^{2}}]}{\beta V(TA)\dot{TA}}]^{\frac{1}{2}}]^{-1}e^{[\frac{d^{2}}{d t^{2}}[\frac{ [4\pi \sqrt{1 +
\dot{TA}^{2} -  TA'^{2}}]}{\beta V(TA)\dot{TA}}]^{\frac{1}{2}}]\sigma} \cosh([\frac{d^{2}}{d t^{2}}[\frac{ [4\pi \sqrt{1 +
\dot{TA}^{2} -  TA'^{2}}]}{\beta V(TA)\dot{TA}}]^{\frac{1}{2}}]\tau)\Big)^{4}}}}
\label{a22}
\end{eqnarray}

The angles $\delta_{I-A}$ and $\delta_{II-A}$ are defined by:

\begin{eqnarray}
&& \cos\delta_{I-A} = \bar{T}_{0,I-A}^{4}\sqrt{1 + \frac{K^{2}}{\Big(a^{-1}e^{-[\frac{d^{2}}{d t^{2}}[\frac{ [4\pi \sqrt{1 +
\dot{TA}^{2} -  TA'^{2}}]}{\beta V(TA)\dot{TA}}]^{\frac{1}{2}}]\sigma} \cosh([\frac{d^{2}}{d t^{2}}[\frac{ [4\pi \sqrt{1 +
\dot{TA}^{2} -  TA'^{2}}]}{\beta V(TA)\dot{TA}}]^{\frac{1}{2}}]\tau)\Big)^{4}}}\nonumber\\&&   \bar{T}_{0,I-A} = \Big(\frac{9\pi^{2}N}{4\sqrt{3}T_{D3}}\Big)^{\frac{1}{2}}T_{0,I-A}
\label{a23}
\end{eqnarray}

\begin{eqnarray}
&& \cos\delta_{II-A} = \bar{T}_{0,II-A}^{4}\sqrt{1 + \frac{K^{2}}{\Big([\frac{d^{2}}{d t^{2}}[\frac{ [4\pi \sqrt{1 +
\dot{TA}^{2} -  TA'^{2}}]}{\beta V(TA)\dot{TA}}]^{\frac{1}{2}}]^{-1}e^{[\frac{d^{2}}{d t^{2}}[\frac{ [4\pi \sqrt{1 +
\dot{TA}^{2} -  TA'^{2}}]}{\beta V(TA)\dot{TA}}]^{\frac{1}{2}}]\sigma} \cosh( [\frac{d^{2}}{d t^{2}}[\frac{ [4\pi \sqrt{1 +
\dot{TA}^{2} -  TA'^{2}}]}{\beta V(TA)\dot{TA}}]^{\frac{1}{2}}]\tau)\Big)^{4}}}\nonumber\\&&   \bar{T}_{0,II-A} = \Big(\frac{9\pi^{2}N}{4\sqrt{3}T_{D3}}\Big)^{\frac{1}{2}}T_{0,II-A}
\label{a24}
\end{eqnarray}
where $T_{0}$ is the temperature of the BIon in non-Rindler space-time. The above equations  show that the metric of the thermal BIon depends on the evolutions of the  tachyonic fields. In fact, the evolution of tachyonic fields has a direct effect on the  thermodynamics of the BIon. Folowing the method in \cite{ss20}, we can obtain the separation distance between two manifolds in a 5-dimensional BIon as:

\begin{eqnarray}
&&  d z_{I-A} = dz_{II-B}\simeq \nonumber\\&& \Big(e^{-4[\frac{d^{2}}{d t^{2}}[\frac{ [4\pi \sqrt{1 +
\dot{TA}^{2} -  TA'^{2}}]}{\beta V(TA)\dot{TA}}]^{\frac{1}{2}}]\sigma}\sinh^{2}( [\frac{d^{2}}{d t^{2}}[\frac{ [4\pi \sqrt{1 +
\dot{TA}^{2} -  TA'^{2}}]}{\beta V(TA)\dot{TA}}]^{\frac{1}{2}}]\tau)\cosh^{2}([\frac{d^{2}}{d t^{2}}[\frac{ [4\pi \sqrt{1 +
\dot{TA}^{2} -  TA'^{2}}]}{\beta V(TA)\dot{TA}}]^{\frac{1}{2}}]\tau)\Big) \times \nonumber\\&& \Big(\frac{F_{DBI,I,A}(\tau,\sigma)\Big(\frac{F_{DBI,I,A}(\tau,\sigma)}{F_{DBI,I,A}(\tau,\sigma_{0})}-e^{-4[\frac{d^{2}}{d t^{2}}[\frac{ [4\pi \sqrt{1 +
\dot{TA}^{2} -  TA'^{2}}]}{\beta V(TA)\dot{TA}}]^{\frac{1}{2}}](\sigma-\sigma_{0})}\frac{\cosh^{2}([\frac{d^{2}}{d t^{2}}[\frac{ [4\pi \sqrt{1 +
\dot{TA}^{2} -  TA'^{2}}]}{\beta V(TA)\dot{TA}}]^{\frac{1}{2}}]\tau_{0})}{\cosh^{2}([\frac{d^{2}}{d t^{2}}[\frac{ [4\pi \sqrt{1 +
\dot{TA}^{2} -  TA'^{2}}]}{\beta V(TA)\dot{TA}}]^{\frac{1}{2}}]\tau)}\Big)^{-\frac{1}{2}}}{F_{DBI,I,A}(\tau_{0},\sigma)\Big(\frac{F_{DBI,I,A}(\tau_{0},\sigma)}{F_{DBI,I,A}(\tau_{0},\sigma_{0})}-e^{-4 [\frac{d^{2}}{d t^{2}}[\frac{ [4\pi \sqrt{1 +
\dot{TA}^{2} -  TA'^{2}}]}{\beta V(TA)\dot{TA}}]^{\frac{1}{2}}](\sigma-\sigma_{0})}\frac{\cosh^{2}( [\frac{d^{2}}{d t^{2}}[\frac{ [4\pi \sqrt{1 +
\dot{TA}^{2} -  TA'^{2}}]}{\beta V(TA)\dot{TA}}]^{\frac{1}{2}}]\tau_{0})}{\cosh^{2}( [\frac{d^{2}}{d t^{2}}[\frac{ [4\pi \sqrt{1 +
\dot{TA}^{2} -  TA'^{2}}]}{\beta V(TA)\dot{TA}}]^{\frac{1}{2}}]\tau)}\Big)^{-\frac{1}{2}}}-\nonumber\\&&\frac{\sinh^{2}( [\frac{d^{2}}{d t^{2}} [\frac{\sqrt{ 1 -2\pi
	l_{s}^{2}G(F)}}{2\pi
	l_{s}^{2} G'(F)F_{01}}]^{\frac{1}{2}}]\tau_{0})}{\sinh^{2}(a\tau)}\Big)^{-\frac{1}{2}}
\label{a25}
\end{eqnarray}

or

\begin{eqnarray}
&&  d z_{I-B} = d z_{II-A}\simeq \nonumber\\&&  \Big(e^{4 [\frac{d^{2}}{d t^{2}}[\frac{ [4\pi \sqrt{1 +
\dot{TA}^{2} -  TA'^{2}}]}{\beta V(TA)\dot{TA}}]^{\frac{1}{2}}]\sigma}\sinh^{2}([\frac{d^{2}}{d t^{2}}[\frac{ [4\pi \sqrt{1 +
\dot{TA}^{2} -  TA'^{2}}]}{\beta V(TA)\dot{TA}}]^{\frac{1}{2}}]\tau)\cosh^{2}( [\frac{d^{2}}{d t^{2}}[\frac{ [4\pi \sqrt{1 +
\dot{TA}^{2} -  TA'^{2}}]}{\beta V(TA)\dot{TA}}]^{\frac{1}{2}}]\tau)\Big) \times \nonumber\\&& \Big(\frac{F_{DBI,II,A}(\tau,\sigma)\Big(\frac{F_{DBI,II,A}(\tau,\sigma)}{F_{DBI,II,A}(\tau,\sigma_{0})}-e^{4[\frac{d^{2}}{d t^{2}}[\frac{ [4\pi \sqrt{1 +
\dot{TA}^{2} -  TA'^{2}}]}{\beta V(TA)\dot{TA}}]^{\frac{1}{2}}](\sigma-\sigma_{0})}\frac{\cosh^{2}( [\frac{d^{2}}{d t^{2}}[\frac{ [4\pi \sqrt{1 +
\dot{TA}^{2} -  TA'^{2}}]}{\beta V(TA)\dot{TA}}]^{\frac{1}{2}}]\tau_{0})}{\cosh^{2}([\frac{d^{2}}{d t^{2}}[\frac{ [4\pi \sqrt{1 +
\dot{TA}^{2} -  TA'^{2}}]}{\beta V(TA)\dot{TA}}]^{\frac{1}{2}}]\tau)}\Big)^{-\frac{1}{2}}}{F_{DBI,II,A}(\tau_{0},\sigma)\Big(\frac{F_{DBI,II,A}(\tau_{0},\sigma)}{F_{DBI,II,A}(\tau_{0},\sigma_{0})}-e^{4 [\frac{d^{2}}{d t^{2}}[\frac{ [4\pi \sqrt{1 +
\dot{TA}^{2} -  TA'^{2}}]}{\beta V(TA)\dot{TA}}]^{\frac{1}{2}}](\sigma-\sigma_{0})}\frac{\cosh^{2}( [\frac{d^{2}}{d t^{2}}[\frac{ [4\pi \sqrt{1 +
\dot{TA}^{2} -  TA'^{2}}]}{\beta V(TA)\dot{TA}}]^{\frac{1}{2}}]\tau_{0})}{\cosh^{2}( [\frac{d^{2}}{d t^{2}}[\frac{ [4\pi \sqrt{1 +
\dot{TA}^{2} -  TA'^{2}}]}{\beta V(TA)\dot{TA}}]^{\frac{1}{2}}]\tau)}\Big)^{-\frac{1}{2}}}-\nonumber\\&&\frac{\sinh^{2}([\frac{d^{2}}{d t^{2}}[\frac{ [4\pi \sqrt{1 +
\dot{TA}^{2} -  TA'^{2}}]}{\beta V(TA)\dot{TA}}]^{\frac{1}{2}}]\tau_{0})}{\sinh^{2}([\frac{d^{2}}{d t^{2}}[\frac{ [4\pi \sqrt{1 +
\dot{TA}^{2} -  TA'^{2}}]}{\beta V(TA)\dot{TA}}]^{\frac{1}{2}}]\tau)}\Big)^{-\frac{1}{2}}  \label{at25}
\end{eqnarray}

with the definition of $F_{DBI,I,A}$ given below:

\begin{eqnarray}
&& F_{DBI,I,A} = F_{DBI,II,B}= \Big( [\frac{d^{2}}{d t^{2}}[\frac{ [4\pi \sqrt{1 +
\dot{TA}^{2} -  TA'^{2}}]}{\beta V(TA)\dot{TA}}]^{\frac{1}{2}}]^{-1}e^{[\frac{d^{2}}{d t^{2}}[\frac{ [4\pi \sqrt{1 +
\dot{TA}^{2} -  TA'^{2}}]}{\beta V(TA)\dot{TA}}]^{\frac{1}{2}}]\sigma}\times \nonumber\\&& \cosh([\frac{d^{2}}{d t^{2}}[\frac{ [4\pi \sqrt{1 +
\dot{TA}^{2} -  TA'^{2}}]}{\beta V(TA)\dot{TA}}]^{\frac{1}{2}}]\tau)\Big)^{2}\frac{4\cosh^{2}\alpha_{I-A} - 3}{\cosh^{4}\alpha_{I-A}} \nonumber\\&&F_{DBI,II,A} = F_{DBI,I,B} =\Big([\frac{d^{2}}{d t^{2}}[\frac{ [4\pi \sqrt{1 +
\dot{TA}^{2} -  TA'^{2}}]}{\beta V(TA)\dot{TA}}]^{\frac{1}{2}}]^{-1}e^{-[\frac{d^{2}}{d t^{2}}[\frac{ [4\pi \sqrt{1 +
\dot{TA}^{2} -  TA'^{2}}]}{\beta V(TA)\dot{TA}}]^{\frac{1}{2}}]\sigma} \times \nonumber\\&&\cosh([\frac{d^{2}}{d t^{2}}[\frac{ [4\pi \sqrt{1 +
\dot{TA}^{2} -  TA'^{2}}]}{\beta V(TA)\dot{TA}}]^{\frac{1}{2}}]\tau)\Big)^{2}\frac{4\cosh^{2}\alpha_{II-A} - 3}{\cosh^{4}\alpha_{II-A}}
\label{a28}
\end{eqnarray}

These separation distances depend on the tachyonic fields and temperature. When the separation distance in one region grows, the separation distance in the other region shrinks. Now, we calculate the area of a thermal BIon by using equations (\ref{a25}),  (\ref{at25}) and (\ref{a10}):

\begin{eqnarray}
&&  A_{I-A} =A_{II-B}= \int \pi r_{I-A}^{2}d z_{I-A}=\int \pi r_{II-B}^{2}d z_{II-B}= \nonumber\\&& \int d\sigma [[\frac{d^{2}}{d t^{2}}[\frac{ [4\pi \sqrt{1 +
\dot{TA}^{2} -  TA'^{2}}]}{\beta V(TA)\dot{TA}}]^{\frac{1}{2}}]^{-1}e^{ [[\frac{d^{2}}{d t^{2}}[\frac{ [4\pi \sqrt{1 +
\dot{TA}^{2} -  TA'^{2}}]}{\beta V(TA)\dot{TA}}]^{\frac{1}{2}}]\sigma} \cosh([\frac{d^{2}}{d t^{2}}[\frac{ [4\pi \sqrt{1 +
\dot{TA}^{2} -  TA'^{2}}]}{\beta V(TA)\dot{TA}}]^{\frac{1}{2}}]\tau) ]\times \nonumber\\&&  \Big(e^{-4[\frac{d^{2}}{d t^{2}}[\frac{ [4\pi \sqrt{1 +
\dot{TA}^{2} -  TA'^{2}}]}{\beta V(TA)\dot{TA}}]^{\frac{1}{2}}]\sigma}\sinh^{2}([\frac{d^{2}}{d t^{2}}[\frac{ [4\pi \sqrt{1 +
\dot{TA}^{2} -  TA'^{2}}]}{\beta V(TA)\dot{TA}}]^{\frac{1}{2}}]\tau)\cosh^{2}([\frac{d^{2}}{d t^{2}}[\frac{ [4\pi \sqrt{1 +
\dot{TA}^{2} -  TA'^{2}}]}{\beta V(TA)\dot{TA}}]^{\frac{1}{2}}]\tau)\Big) \times \nonumber\\&& \Big(\frac{F_{DBI,I,A}(\tau,\sigma)\Big(\frac{F_{DBI,I,A}(\tau,\sigma)}{F_{DBI,I,A}(\tau,\sigma_{0})}-e^{-4[\frac{d^{2}}{d t^{2}}[\frac{ [4\pi \sqrt{1 +
\dot{TA}^{2} -  TA'^{2}}]}{\beta V(TA)\dot{TA}}]^{\frac{1}{2}}](\sigma-\sigma_{0})}\frac{\cosh^{2}([\frac{d^{2}}{d t^{2}}[\frac{ [4\pi \sqrt{1 +
\dot{TA}^{2} -  TA'^{2}}]}{\beta V(TA)\dot{TA}}]^{\frac{1}{2}}]\tau_{0})}{\cosh^{2}( [\frac{d^{2}}{d t^{2}} [\frac{\sqrt{ 1 -2\pi
	l_{s}^{2}G(F)}}{2\pi
	l_{s}^{2} G'(F)F_{01}}]^{\frac{1}{2}}]\tau)}\Big)^{-\frac{1}{2}}}{F_{DBI,I,A}(\tau_{0},\sigma)\Big(\frac{F_{DBI,I,A}(\tau_{0},\sigma)}{F_{DBI,I,A}(\tau_{0},\sigma_{0})}-e^{-4[\frac{d^{2}}{d t^{2}}[\frac{ [4\pi \sqrt{1 +
\dot{TA}^{2} -  TA'^{2}}]}{\beta V(TA)\dot{TA}}]^{\frac{1}{2}}](\sigma-\sigma_{0})}\frac{\cosh^{2}([\frac{d^{2}}{d t^{2}}[\frac{ [4\pi \sqrt{1 +
\dot{TA}^{2} -  TA'^{2}}]}{\beta V(TA)\dot{TA}}]^{\frac{1}{2}}]\tau_{0})}{\cosh^{2}([\frac{d^{2}}{d t^{2}}[\frac{ [4\pi \sqrt{1 +
\dot{TA}^{2} -  TA'^{2}}]}{\beta V(TA)\dot{TA}}]^{\frac{1}{2}}]\tau)}\Big)^{-\frac{1}{2}}}-\nonumber\\&&\frac{\sinh^{2}([\frac{d^{2}}{d t^{2}}[\frac{ [4\pi \sqrt{1 +
\dot{TA}^{2} -  TA'^{2}}]}{\beta V(TA)\dot{TA}}]^{\frac{1}{2}}]\tau_{0})}{\sinh^{2}([\frac{d^{2}}{d t^{2}}[\frac{ [4\pi \sqrt{1 +
\dot{TA}^{2} -  TA'^{2}}]}{\beta V(TA)\dot{TA}}]^{\frac{1}{2}}]\tau)}\Big)^{-\frac{1}{2}}
\label{ay25}
\end{eqnarray}

or

\begin{eqnarray}
&& A_{II-A} =A_{I-B}= \int \pi r_{II-A}^{2}d z_{II-A}=\int \pi r_{I-B}^{2}d z_{I-B}= \nonumber\\&& \int d\sigma [[\frac{d^{2}}{d t^{2}}[\frac{ [4\pi \sqrt{1 +
\dot{TA}^{2} -  TA'^{2}}]}{\beta V(TA)\dot{TA}}]^{\frac{1}{2}}]^{-1}e^{ -[\frac{d^{2}}{d t^{2}}[\frac{ [4\pi \sqrt{1 +
\dot{TA}^{2} -  TA'^{2}}]}{\beta V(TA)\dot{TA}}]^{\frac{1}{2}}]\sigma} \cosh([\frac{d^{2}}{d t^{2}}[\frac{ [4\pi \sqrt{1 +
\dot{TA}^{2} -  TA'^{2}}]}{\beta V(TA)\dot{TA}}]^{\frac{1}{2}}]\tau) ]\times \nonumber\\&& \Big(e^{4[\frac{d^{2}}{d t^{2}}[\frac{ [4\pi \sqrt{1 +
\dot{TA}^{2} -  TA'^{2}}]}{\beta V(TA)\dot{TA}}]^{\frac{1}{2}}]\sigma}\sinh^{2}( [\frac{d^{2}}{d t^{2}}[\frac{ [4\pi \sqrt{1 +
\dot{TA}^{2} -  TA'^{2}}]}{\beta V(TA)\dot{TA}}]^{\frac{1}{2}}]\tau)\cosh^{2}( [[\frac{d^{2}}{d t^{2}}[\frac{ [4\pi \sqrt{1 +
\dot{TA}^{2} -  TA'^{2}}]}{\beta V(TA)\dot{TA}}]^{\frac{1}{2}}]\tau)\Big) \times \nonumber\\&& \Big(\frac{F_{DBI,II,A}(\tau,\sigma)\Big(\frac{F_{DBI,II,A}(\tau,\sigma)}{F_{DBI,II,A}(\tau,\sigma_{0})}-e^{4 [\frac{d^{2}}{d t^{2}}[\frac{ [4\pi \sqrt{1 +
\dot{TA}^{2} -  TA'^{2}}]}{\beta V(TA)\dot{TA}}]^{\frac{1}{2}}](\sigma-\sigma_{0})}\frac{[\frac{d^{2}}{d t^{2}}[\frac{ [4\pi \sqrt{1 +
\dot{TA}^{2} -  TA'^{2}}]}{\beta V(TA)\dot{TA}}]^{\frac{1}{2}}]\tau_{0})}{\cosh^{2}([\frac{d^{2}}{d t^{2}}[\frac{ [4\pi \sqrt{1 +
\dot{TA}^{2} -  TA'^{2}}]}{\beta V(TA)\dot{TA}}]^{\frac{1}{2}}]\tau)}\Big)^{-\frac{1}{2}}}{F_{DBI,II,A}(\tau_{0},\sigma)\Big(\frac{F_{DBI,II,A}(\tau_{0},\sigma)}{F_{DBI,II,A}(\tau_{0},\sigma_{0})}-e^{4 [\frac{d^{2}}{d t^{2}}[\frac{ [4\pi \sqrt{1 +
\dot{TA}^{2} -  TA'^{2}}]}{\beta V(TA)\dot{TA}}]^{\frac{1}{2}}](\sigma-\sigma_{0})}\frac{\cosh^{2}( [\frac{d^{2}}{d t^{2}}[\frac{ [4\pi \sqrt{1 +
\dot{TA}^{2} -  TA'^{2}}]}{\beta V(TA)\dot{TA}}]^{\frac{1}{2}}]\tau_{0})}{\cosh^{2}([\frac{d^{2}}{d t^{2}}[\frac{ [4\pi \sqrt{1 +
\dot{TA}^{2} -  TA'^{2}}]}{\beta V(TA)\dot{TA}}]^{\frac{1}{2}}]\tau)}\Big)^{-\frac{1}{2}}}-\nonumber\\&&\frac{\sinh^{2}([\frac{d^{2}}{d t^{2}}[\frac{ [4\pi \sqrt{1 +
\dot{TA}^{2} -  TA'^{2}}]}{\beta V(TA)\dot{TA}}]^{\frac{1}{2}}]\tau_{0})}{\sinh^{2}([\frac{d^{2}}{d t^{2}}[\frac{ [4\pi \sqrt{1 +
\dot{TA}^{2} -  TA'^{2}}]}{\beta V(TA)\dot{TA}}]^{\frac{1}{2}}]\tau)}\Big)^{-\frac{1}{2}}  \label{ayt25}
\end{eqnarray}

The above equations shows that the area of the thermal accelerating BIons depends on the tachyonic fields which live on them. These  fields lead to the acceleration of the BIon. This acceleration produces a Rindler space-time with two regions. By increasing the strength of the tachyonic fields, the area of a BIon in region I expands, while the area of a BIon in region II contracts.

\section{ Dependency of the Tsallis entropy of a BIon on tachyons}\label{o2}

In this section, we will consider the effect of tachyonic fields on the entropy of the BIon. We will show that tachyonic fields lead to the expansion of the BIon and increasing the  entropy of the BIon in one region and decreasing the entropy in the other region. Previously, thermodynamical parameters like the entropy have been obtained in \cite{ss13,ss14,ss15,ss20}. Using those relations and replacing the acceleration by tachyonic fields in equation (\ref{a10}), we obtain:

\begin{eqnarray}
&& d S_{I-A}=d S_{II-B}=\frac{4 T_{D3}^{2}}{\pi T_{0,I-A}^{5}}F_{DBI,I,A}(\sigma,\tau) \Big(\frac{1}{[\frac{d^{2}}{d t^{2}}[\frac{ [4\pi \sqrt{1 +
\dot{TA}^{2} -  TA'^{2}}]}{\beta V(TA)\dot{TA}}]^{\frac{1}{2}}]}e^{ [\frac{d^{2}}{d t^{2}}[\frac{ [4\pi \sqrt{1 +
\dot{TA}^{2} -  TA'^{2}}]}{\beta V(TA)\dot{TA}}]^{\frac{1}{2}}]\sigma} \cosh(a\tau)\Big)^{2} \times \nonumber\\&&  \frac{\Big(\sinh^{2}( [\frac{d^{2}}{d t^{2}}[\frac{ [4\pi \sqrt{1 +
\dot{TA}^{2} -  TA'^{2}}]}{\beta V(TA)\dot{TA}}]^{\frac{1}{2}}]\tau)+cosh^{2}( [\frac{d^{2}}{d t^{2}}[\frac{ [4\pi \sqrt{1 +
\dot{TA}^{2} -  TA'^{2}}]}{\beta V(TA)\dot{TA}}]^{\frac{1}{2}}]\tau)\Big)}{\sqrt{F_{DBI,I,A}^{2}(\sigma,\tau)-F_{DBI,I,A}^{2}(\sigma_{o},\tau)}}\times \nonumber\\&& \frac{4 }{\cosh^{4}\alpha_{I-A}}
\label{a26}
\end{eqnarray}

and

\begin{eqnarray}
&& d S_{II-A}= d S_{I-B}=\frac{4 T_{D3}^{2}}{\pi T_{0,II-A}^{5}}F_{DBI,II,A}(\sigma,\tau) \Big(\frac{1}{[\frac{d^{2}}{d t^{2}}[\frac{ [4\pi \sqrt{1 +
\dot{TA}^{2} -  TA'^{2}}]}{\beta V(TA)\dot{TA}}]^{\frac{1}{2}}]}e^{ -[\frac{d^{2}}{d t^{2}}[\frac{ [4\pi \sqrt{1 +
\dot{TA}^{2} -  TA'^{2}}]}{\beta V(TA)\dot{TA}}]^{\frac{1}{2}}]\sigma} \cosh(a\tau)\Big)^{2} \times \nonumber\\&&  \frac{\Big(\sinh^{2}( [\frac{d^{2}}{d t^{2}}[\frac{ [4\pi \sqrt{1 +
\dot{TA}^{2} -  TA'^{2}}]}{\beta V(TA)\dot{TA}}]^{\frac{1}{2}}]\tau)+cosh^{2}( [\frac{d^{2}}{d t^{2}}[\frac{ [4\pi \sqrt{1 +
\dot{TA}^{2} -  TA'^{2}}]}{\beta V(TA)\dot{TA}}]^{\frac{1}{2}}]\tau)\Big)}{\sqrt{F_{DBI,II,A}^{2}(\sigma,\tau)-F_{DBI,II,A}^{2}(\sigma_{o},\tau)}}\times \nonumber\\&& \frac{4 }{\cosh^{4}\alpha_{II-A}}\label{a27}
\end{eqnarray}

The above equations show that the entropy of a thermal BIon depends on tachyonic fields which live on it. These fields lead to the acceleration of a thermal BIon. Due to this acceleration, a Rindler horizon  emerges and two regions appear. The entropy of the BIon in one region is opposite to that in another region. This means that by increasing the entropy of the BIon in one region, the entropy of a BIon in another region decreases. Also, the entropy of the BIon in one end is opposite to the entropy of the BIon in the other end in that region. Consequently, the entropy of brane A in region I acts oppositely  to the entropy of brane A in region II, and also acts similar to the entropy  of brane B in region II.

Now, comparing the entropies in equations (\ref{ay25}) and (\ref{ayt25}) with the areas in equations (\ref{a26}) and (\ref{a27}), we  obtain the relation between entropy and area:

\begin{eqnarray}
&&  S_{I-A} = S_{II-B}= [1+[\frac{4 T_{D3}^{2}}{\pi T_{0,I-A}^{5}}]^{-1}tanh^{2}( [\frac{d^{2}}{d t^{2}}[\frac{ [4\pi \sqrt{1 +
\dot{TA}^{2} -  TA'^{2}}]}{\beta V(TA)\dot{TA}}]^{\frac{1}{2}}]\tau)]^{-\frac{2}{5}}A^{\frac{3}{8}}\label{aq1}
\end{eqnarray}

\begin{eqnarray}
&&  S_{II-A} = S_{I-B}= [1+[\frac{4 T_{D3}^{2}}{\pi T_{0,II-A}^{5}}]^{-1}coth^{2}( [\frac{d^{2}}{d t^{2}}[\frac{ [4\pi \sqrt{1 +
\dot{TA}^{2} -  TA'^{2}}]}{\beta V(TA)\dot{TA}}]^{\frac{1}{2}}]\tau)]^{\frac{4}{7}}A^{-\frac{5}{4}}\label{aq1}
\end{eqnarray}

The above equations show that entropies have direct relations with the areas of a BIon. This is in good agreement with the prediction of Tsallis whereby entropy should have the form  $S = \gamma A^{\beta}$.   These entropies depend on the tachyonic potential and temperature of the system.  In one region, by increasing the area, the entropy increases, while in another region by increasing the area, the entropy decreases. In fact, thermodynamically, the behaviour of the BIon in each region is opposite to the behaviour of a BIon in another region.

\section{Summary} \label{sum}

In this research, we have shown that the tachyonic potential between branes and anti-branes leads to the emergence of a wormhole between them. This wormhole and two branes form a tachyonic BIon.  In this type of BIon, the tachyonic potential leads to the acceleration of branes and emergence of a Rindler space-time. This space-time includes two regions in each of which a tachyonic BIon lives. We have shown that the entropy of these BIons includes some terms similar to the Tsallis entropy and could be
written as $S = \gamma A^{\beta}$, where A is the area of the BIon. By increasing the  tachyonic potential, the entropy of a BIon in one region increases, while the entropy of a BIon in another region decreases.

\section*{Acknowledgements}
\noindent The authors have no conflict of interest. Each author has contributed equally.  The work of A. Sepehri has been supported financially by the Research Institute for Astronomy-Astrophysics of Maragha (RIAAM) under research project number 1/5750-20.

\end{document}